\documentclass[aps,prl,reprint,preprintnumbers,showpacs,showkeys,superscriptaddress]{revtex4-1}
\usepackage{amsmath,amssymb,bm,mathrsfs}
\usepackage{srcltx}
\usepackage[colorlinks=true,citecolor=blue,linkcolor=blue]{hyperref}

\begin{document}
\title{Noncommuting Momenta of Topological Solitons}

\author{Haruki Watanabe}
\email{hwatanabe@berkeley.edu}
\affiliation{Department of Physics, University of California,
  Berkeley, California 94720, USA}

\author{Hitoshi Murayama}
\email{hitoshi@berkeley.edu, hitoshi.murayama@ipmu.jp}
\affiliation{Department of Physics, University of California,
  Berkeley, California 94720, USA} 
\affiliation{Theoretical Physics Group, Lawrence Berkeley National
  Laboratory, Berkeley, California 94720, USA} 
\affiliation{Kavli Institute for the Physics and Mathematics of the
  Universe (WPI), Todai Institutes for Advanced Study, University of Tokyo,
  Kashiwa 277-8583, Japan} 

\begin{abstract}
We show that momentum operators of a topological soliton may not commute among themselves when the soliton is associated with the second cohomology $H^2$ of the target space.  The commutation relation is proportional to the winding number,  taking a constant value within each topological sector. The noncommutativity makes it impossible to specify the momentum of a topological soliton, and induces a Magnus force.  
\end{abstract}
\preprint{UCB-PTH 14/02, IPMU14-0019}
\maketitle

\paragraph{Introduction.}
---When quantum mechanics was discovered, its most surprising aspect was the uncertainty principle $\Delta x \Delta p \geq \hbar/2$, namely, that one cannot specify both the position and momentum of a particle at the same time.  It originates from the canonical commutation relation $[x^i, p^j] = i \hbar \delta^{ij} \neq 0$.  Yet, it has been taken for granted that any object can move with a definite momentum in the absence of an external field.  It is mathematically possible, however, that the algebra of the momentum operators has a {\it central extension}\/
\begin{eqnarray}
[P^i,P^j]=i\hbar C^{ij},
\end{eqnarray}
where $C^{ij}$'s commute with all other symmetry generators.  

We show in this Letter that the momentum operators do not commute for certain topological solitons in quantum field theories without an external field.  Thus, one cannot fully specify the momentum of the object, because different components are subject to uncertainty relations.  The commutator $C_{ij} = \epsilon_{ij} N$ turns out to be given by the topological number $N$ and hence is stable against perturbations.  This commutation relation has not been discussed in the literature, nor has its topological nature.  

There are two reasons why extensions do not usually appear in quantum field theories. First, since the momentum density $T^{0i}(\vec{x},t)$ transforms as a scalar under translation, it should obey~\cite{Weinberg}
\begin{eqnarray}
(i\hbar)^{-1}[P^i,T^{0j}(\vec{x},t)]=-\nabla_iT^{0j}(\vec{x},t).\label{PT1}
\end{eqnarray}
By taking the volume integration of both sides of the equation, we see that $P^i$ and $P^j$ do commute, {\it provided that $T^{0j}(\vec{x},t)\rightarrow 0$ as $|\vec{x}|\rightarrow\infty$}\/.  We note that Eq.~\eqref{PT1} assumes {\it the absence of singularities}\/, as we shall see below. Therefore, in order to have a nonzero extension, we need to allow finite contribution from the spatial boundary of the system or singularities in fields.

Another reason why the momentum operators should commute is due to the rotational or boost symmetry~\cite{Weinberg}.  If the system is fully invariant under spatial rotation, the angular momentum $M^{ij}$ ($i,j=1,\ldots,d$ and $d$ is the spatial dimensions) joins the algebra of the symmetry generators. Then, the Jacobi identity among $P^i$, $P^j$, and $M^{ik}$ ($i\neq j$, $k\neq i,j$) demands $[P^i,P^j]=i\hbar C^{ij}=0$.  (Note that the inplane rotation $M^{ij}$ does not affect $C^{ij}$.  For example, when $d=2$, $[P^x,P^y]$ can be nonzero even in the presence of rotational symmetry around the $z$ axis.)  Similarly, the Jacobi identity among the boost operator $B^i$, the momentum $P^j$, and the Hamiltonian $H$, with the help of $[H,B^i]=i\hbar P^i$, leads to $C^{ij}=0$.  Here, $B^i$ can be either the Lorentz boost or the Galilean boost.

An external uniform magnetic field $\vec{B}$ makes it possible to go around these two obstacles.  First, the Lagrangian does not have naive translational symmetry due to the explicit coordinate dependence of the vector potential; the definition of the conserved momentum operator must be modified, replacing Eq.~\eqref{PT1} with
\begin{eqnarray}
(i\hbar)^{-1}[P_B^i,T_B^{0j}]=-\nabla_iT_B^{0j}-q_a\epsilon^{ijk}B^k j_a^0,\label{PT2}
\end{eqnarray}
where $\vec{P}_B$ is the conserved momentum under the uniform magnetic field, the so-called magnetic translational operator, $T_B^{0i}(\vec{x},t)$ denotes its density, $q_a$ is the charge of particles of a certain kind labeled by $a$, and $j_a^0(\vec{x},t)$ is their number density~\footnote{Eq.~\eqref{PT2} assumes that fields are smooth and the density $j_a^0(\vec{x},t)$ vanishes fast enough as $|\vec{x}|\rightarrow\infty$.}.  Moreover, the magnetic field breaks the boost symmetry and the spatial rotation down to the rotation around the axis of the magnetic field.  As a result, nonzero extensions are allowed in the plane perpendicular to the magnetic field and, by integrating Eq.~\eqref{PT2}, we indeed obtain $C^{i j}=-q_a\epsilon^{ijk}B^k N_a$ with $N_a=\int\mathrm{d}^dx\,j_a^0(\vec{x},t)$.

In this Letter, we show that topological solitons corresponding to the second cohomology $H^2(G/H)$ of the target space $G/H$ allow for nonzero extensions without an external field.  The first obstacle is overcome by the presence of singularities or boundary contributions.  The second one is evaded in field theories without rotation or boost symmetries.  We also discuss the central extensions for the topological defects characterized by $\pi_1(G/H)$, such as vortices in superfluids.

The central extensions among components of the momentum operator have many important physical consequences.   Topological solitons in general spontaneously break the translational symmetry.  In two space dimensions, they are pointlike objects and their center of mass motion does not cost energy.  The corresponding zero modes (translational moduli) may be seen as Nambu-Goldstone modes of the broken translations and it is mandatory to discuss their quantum mechanics~\cite{Shifman}.  As we shall see below, nonzero central extensions drastically affect the dynamics of these low-energy modes by reducing the number of independent modes~\cite{Nambu2,WatanabeMurayama1,Hidaka} and inducing the Magnus force on them~\cite{Thouless}.

The fact that spacetime symmetries can be centrally extended by a topological charge is reminiscent of field theories with extended supersymmetry, where the anticommutator $\{Q_i, Q_j\} = \epsilon_{ij} Z$ of supercharges is proportional to the magnetic charge \cite{Witten:1978mh}, or with noncommutative geometry $[x_\mu, x_\nu] \neq 0$ that appears in string theory compactifications \cite{Seiberg:1999vs}.

\paragraph{Skyrmions in ferromagnets.}---
To illustrate our main result in the simplest example, let us consider Skyrmions in ferromagnets in $2+1$ dimensions.  Skyrmions are continuous spin configurations $\vec{n}(\vec{x},t)$ ($\vec{n}^2=1$) with a nontrivial winding $\pi_2(S^2)=\mathbb{Z}$.  Here we consider a smooth spin configuration with the boundary condition $\vec{n}(\vec{x},t)\rightarrow(0,0,1)^T$ as $|\vec{x}|\rightarrow\infty$.  The effective Lagrangian for Heisenberg ferromagnets in the continuum limit reads
\begin{equation}
\mathcal{L}=s\dot{\phi}(\cos\theta-1)-(J/2)\nabla_i\vec{n}\cdot\nabla_i\vec{n},\label{ferro}
\end{equation}
where $s=S/a^2$ is the magnetization density ($S$ is the spin of each variable on a microscopic lattice and $a$ is the lattice constant), $(\theta,\phi)$ is the spherical coordinate of the unit vector $\vec{n}$.  The first term in the Lagrangian, known as the Berry phase term, implies the canonical commutation relation $[\phi(\vec{x},t),p_\phi(\vec{x}',t)]=i\hbar \delta^2(\vec{x}-\vec{x}')$ with $p_\phi=s(\cos\theta-1)$.  Note that we could well have chosen $s\dot{\phi}(\cos\theta+1)$ for the first term, which differs only by a total time derivative.

The Lagrangian \eqref{ferro} does not possess the boost symmetry and has only rotational symmetry around the $z$ axis, so the extension is allowed.  We show that $C^{xy}=4\pi s N$, where $N$ is the winding number of $\pi_2(S^2)$.

Applying Noether's theorem to the translational symmetry $\vec{n}'(\vec{x}+\vec{\epsilon},t)=\vec{n}(\vec{x},t)$, one can derive the momentum density $T^{0i}(\vec{x},t)=p_\phi(\vec{x},t)\nabla_i\phi(\vec{x},t)$ and the momentum operator $P^i=\int\mathrm{d}^2x\,T^{0i}$.  Note that we chose the surface term in Eq.~\eqref{ferro} in such a way that the momentum density $T^{0i}$ vanishes as $|\vec{x}|\rightarrow \infty$.  Although the spin texture $\vec{n}$ is smooth, the field $\phi$ does have a vortex singularity at $\theta=\pi$ in the presence of Skyrmions.  From a straightforward calculation, we find that Eq.~\eqref{PT1} receives an additional contribution from the singularity,
\begin{eqnarray}
(i\hbar)^{-1}[P^x,T^{0y}]=-\nabla_xT^{0y}+p_\phi\epsilon^{ij}\nabla_i\nabla_j\phi. 
\label{PT3}
\end{eqnarray}
By taking the volume integration and integrating by parts,
\begin{eqnarray}
(i\hbar)^{-1}[P^x,P^y]&=& s\int\mathrm{d}^2x\sin\theta\epsilon^{ij}\nabla_i\theta\nabla_j\phi=4\pi sN,\label{extension1}
\end{eqnarray}
where $N\equiv(4\pi)^{-1}\int\mathrm{d}^2x\,\vec{n}\cdot\nabla_x\vec{n}\times\nabla_y\vec{n}\in\mathbb{Z}$ is the winding number of the configuration $\vec{n}(\vec{x},t)$.  If we had used the Lagrangian $s\dot{\phi}(\cos\theta+1)$ instead, the singularity at the center of the Skyrmion is canceled by $p_\phi = s (\cos\theta+1)=0$, while the surface term $-\int d^2 x \epsilon_{ij} \nabla_iT^{0j}$ leads to the same result.  The spin algebra $[S^i,S^j]=i\hbar\epsilon^{ijk}S^k$ and the commutation relation between the spin and the momentum $[P^i,S^j]=0$ are not modified, even in the presence of Skyrmions.

Let us now discuss the effect of the central extension on the low-energy dynamics of a Skyrmion.  Using the stereographic projection $\vec{n}=(1+\bar{z}z)^{-1}\left(\bar{z}+z,i(\bar{z}-z),\bar{z}z-1\right)^T$, one finds $z=\rho_0^{-1} [(x-x_0)+i(y-y_0)]e^{i\theta_0}$ is a solution to the equation of motion that is consistent with the boundary condition and has the unit winding number.  Here, $x_0,y_0,\rho_0,\theta_0$ are constants and the energy of the configuration does not depend on them. They are called moduli of the soliton; the low-energy dynamics of the Skyrmion can be described by allowing time dependence to them~\cite{Shifman}.  Substituting the solution into the Lagrangian~\eqref{ferro} and performing the spatial integration, we find
\begin{eqnarray}
L_{\text{eff}}&=&\frac{C^{xy}}{2}(y_0\dot{x}_0-x_0\dot{y}_0)-\Delta_z(\rho_0)\dot{\theta}_0\notag\\
&&+\frac{m}{2}(\dot{x}_0^2+\dot{y}_0^2)+\frac{m'(\rho_0)}{2}\dot{\rho}_0^2+\frac{m''(\rho_0)}{2}\dot{\theta}_0^2\label{skeff}
\end{eqnarray}
to the quadratic order in the time derivative, where $\Delta_z=s\int\mathrm{d}^dx\left[n_z(\vec{x})-1\right]$ measures the difference of the $z$ component of the spin from the ground state.  The second line of Eq.~\eqref{skeff} is generated by integrating out gapped degrees of freedom; $m$ is the mass of the Skyrmion and $m',m''$ are functions of $\rho_0$~\footnote{$\Delta_z,m',m''$ diverge logarithmically at large radii for a single Skyrmion, while are finite for multi-Skyrmions.  One can avoid their divergences by, {\it e.g.}\/, applying a small external magnetic field $\mu s B^z n^z$, which sets the size of Skyrmion $\lambda=(\mu S B^z/J)^{-1/2}a$, or by considering a finite-size system.  A Hopf term, if present, gives an additional contribution $\hbar\Theta/2\pi$ to $\Delta_z$~\cite{UJW}.}.   See Ref.~\cite{UJW}, which discusses Skyrmions in {\it antiferromagnets}\/ (that do not produce $C^{xy}$), for more details.

The central extension drastically affects the dynamics of the translational moduli.  First, it reduces the number of independent degrees of freedom in the low-energy limit.  Indeed, if we neglect the $O(\partial_t^2)$ term, $x_0$ and $y_0$ become canonically conjugate to each other due to the first term in Eq.~\eqref{skeff}.  Second, the Skyrmion feels the Magnus force.  To see this, we compare our effective Lagrangian to the single-particle Lagrangian under an uniform magnetic field $m(\dot{x}^2+\dot{y}^2)/2+qB(x\dot{y}-y\dot{x})/2$.  We immediately notice that the center of mass motion of the Skyrmion is identical to the single-particle motion in a magnetic field $qB=-C^{xy}$.  The Lorentz force $\vec{F}=q\dot{\vec{x}}_0\times B\hat{z}$ acting on the Skyrmion is nothing but the Magnus force~\cite{Volovik,Stone}.  

When we quantize the translational moduli of a soliton, we normally expect it to be in an eigenstate of the momentum, namely a plane wave.  However, because of the noncommuting momenta, we cannot specify $P^x$ and $P^y$ of the soliton at the same time, and the soliton is in a Landau-level wave function.  The degeneracy of the ground state is given by $N_{\text{deg}}=|qB|A/(2\pi \hbar)=2sA/\hbar$ ($A$ is the area of the system), which is indeed an integer as $sA$ represents the total spin in the system.  

In three spatial dimensions, we can consider a Skyrmion line configuration and discuss its vibration.  In this case the effective Lagrangian reads $-C^{xy}\epsilon_{ij}x_0^i\dot{x}_0^j/2-g(\nabla_z\vec{x}_0)^2/2+\cdots$, describing a single mode with a quadratic dispersion $\omega=(g/C^{xy})k_z^2$.  

Finally, Skyrmions may spontaneously form a lattice 
in two spatial dimensions, as predicted theoretically~\cite{1989,2002,2006,Ashvin1,Ashvin2,Tewari,Fischer} and realized experimentally \cite{Pfleiderer2009c,Pfleiderer2010,Tokura2010FeGe,Tokura2010FeCoSi,IronSkX,Adams,Seki}.  
Suppose that there are $N_{\text{sk}}$ Skyrmions in the system and each of them has the winding number $Q$. Then, the effective magnetic field acting on each Skyrmion is $qB=C^{xy}/N_{\text{sk}}=4\pi s Q$ and the area of the unit cell is $A_{\text{uc}}=A/N_{\text{sk}}$. In this setup, the filling factor $\nu=N_{\text{sk}}/N_{\text{deg}}=(\hbar/2S)(a^2/A_{\text{uc}})Q^{-1}$ must be quite small, since we need $a^2\ll A_{\text{uc}}$ to verify the continuum description of spins. Therefore, Skyrmions tend to be in a crystal phase, rather than a melted quantum Hall state.  The low-energy effective Lagrangian for phonons is given by $\mathcal{L}_{\text{eff}}=-(C^{xy}/2A)\epsilon_{ij}x_0^i\dot{x}_0^j-(g^{ij}_{k\ell}/2)\nabla_i x_0^k\nabla_j x_0^\ell$, which describes a single phonon branch with a quadratic dispersion~\cite{Nagaosa,Oleg}.  We can understand all of these phenomena on the same footing without detailed numerical simulations.

\paragraph{General coset space.}---
Let us now turn to more general cases.  The Lagrangian that describes the low-energy dynamics of Nambu-Goldstone bosons is the nonlinear sigma model with the target space $G/H$.  We assume that $G/H$ is compact and work in $2+1$ dimensions for simplicity.  Let $\pi^a$ ($a=1,\ldots,\text{dim}\,G/H$) be a local coordinate of $G/H$. The set of the fields $\pi^a(\vec{x},t)$ is a map $\pi:\mathbb{R}^2\times \mathbb{R}\rightarrow G/H$.

In general, the time-derivative terms of the effective Lagrangian take the form
\begin{equation}
\mathcal{L}_{\text{eff}}^{\text{(time)}}=c_a(\pi)\dot{\pi}^a+(1/2)g_{ab}(\pi)\dot{\pi}^a\dot{\pi}^b\label{eq:Ltime}
\end{equation}
to the quadratic order in time derivatives~\cite{Leutwyler:1994nonrel, WatanabeMurayama1, WatanabeMurayama3}.  The explicit formula for $c_a$ and $g_{ab}$ in terms of the Maurer-Cartan form are given in Refs.~\cite{WatanabeMurayamaBrauner1,WatanabeMurayama3}.  The first term is identified as Berry's phase~\cite{WatanabeMurayama3}. It defines a (partially) symplectic structure on $G/H$, making some fields canonically conjugate to each other in the low-energy limit~\cite{WatanabeMurayama1}. 

When the second cohomology $H^2(G/H)$ is nontrivial, the first term of the effective Lagrangian may be seen as the Wess-Zumino term by regarding the time axis as a closed loop $S^1$ on $G/H$.  In such a case, the two-form
\begin{equation}
\omega\equiv\mathrm{d}[c_a(\pi)\mathrm{d}\pi^a]=\frac{1}{2}\left[\frac{\partial c_b(\pi)}{\partial \pi^a}-\frac{\partial c_a(\pi)}{\partial \pi^b}\right]\mathrm{d}\pi^a\wedge\mathrm{d}\pi^b\label{twoform}
\end{equation}
represents a nontrivial element of $H^2(G/H)$ ($c_a\mathrm{d}\pi^a$ is not necessarily defined globally on $G/H$).  There exists a two-cycle $C_2$ of $G/H$ such that $C_2$ is not a two-boundary and $2\pi \hbar n_0\equiv\int_{C_2}\omega\neq0$~\cite{Nakahara}.  Note that $n_0A$ must also be quantized to an integer so that the action is well defined modulo $2\pi\hbar$ \cite{Witten:1983tw}.  For example, in the case of ferromagnets, $G/H=S^2$ and $H^2(S^2)=\mathbb{Z}$ is nontrivial.  From Eq.~\eqref{ferro}, one can read $c_\theta(\theta,\phi)=0$ and $c_\phi(\theta,\phi)=s(\cos\theta-1)$, $\int_{S^2}\omega=-s\int_{S^2}\sin\theta\mathrm{d}\theta\wedge\mathrm{d}\phi=-4\pi s$, and $n_0A=-2sA/\hbar\in\mathbb{Z}$.

We choose $\pi^a(\vec{x},t)=0$ as our ground state at work and we consider a configuration that approaches to it as $|\vec{x}|\rightarrow\infty$.  Thus, we effectively compactify our space manifold $\mathbb{R}^2$ into $S^2$.  One can set $c_a(0)=0$ without loss of generality since $c_a(0)\dot{\pi}^a$ is a surface term.  This is a way to eliminate contribution to the momentum density from the spatial infinity.

In Eq.~(\ref{eq:Ltime}), $\mathcal{L}_{\text{eff}}^{\text{(time)}}$ defines the canonical conjugate to $\pi^a$, $p_a=\partial \mathcal{L}_{\text{eff}}/\partial\dot{\pi}^a=g_{ab}\dot{\pi}^b+c_a$. Assuming that $g_{ab}$ is nondegenerate, one can set up the canonical commutation relation $[\pi^a(\vec{x},t),p_b(\vec{x}',t)]=i\hbar\delta^a_b\delta^2(\vec{x}-\vec{x}')$~\footnote{The Lagrangian is dominated by the $O(\partial_t)$ term in the low-energy limit, and we can neglect the $O(\partial_t^2)$ term for type-B Nambu-Goldstone bosons. If we do so, however, the Lagrangian is constrained and we should follow Dirac's quantization procedure.  The end result, of course, is unchanged.}.  The commutation relation among the momentum operator $P^i=\int \mathrm{d}^2x\,T^{0i}=\int \mathrm{d}^2x\,p_a\nabla_i\pi^a$ is given by
\begin{eqnarray}
&&(i\hbar)^{-1}[P^x,P^y]=-\int\mathrm{d}^2x\,\epsilon^{ij}\nabla_ip_a\nabla_j\pi^a.\label{comm}
\end{eqnarray}
For a static field configuration, $p_a=c_a(\pi)$ and Eq.~\eqref{comm} reduces to
\begin{eqnarray}
(i\hbar)^{-1}[P^x,P^y]&=&-\int\mathrm{d}^2x\left(\frac{\partial c_b}{\partial \pi^a}-\frac{\partial c_a}{\partial \pi^b}\right)\frac{\epsilon^{ij}}{2}\nabla_i\pi^a\nabla_j\pi^b\nonumber\\
&=&-\int\pi^*\omega=-2\pi \hbar n_0N.\label{general}
\end{eqnarray}
Here, $\pi^*\omega$ represents the pull back of the two-form $\omega\in H^2(G/H)$ by the static map $\pi: \mathbb{R}^2\rightarrow G/H$ and $N$ is the winding number of the map around the two-cycle $C_2$.  If $H^2(G/H)$ is trivial, the integral in the first line of Eq.~\eqref{general} vanishes because the two-form in Eq.~\eqref{twoform} is exact and all two-cycles of $G/H$ are two-boundaries.  The derivation of Eqs.~\eqref{comm} and \eqref{general} assumes that $p_a\rightarrow 0$ as $|\vec{x}|\rightarrow 0$, but the end result does not depend on the constant part of $p_a$; they are valid more generally in the following analysis.

We would like to see explicitly that the momentum operators are indeed those of the soliton. The Lagrangian density can be rewritten as $\mathcal{L}=p_a\dot{\pi}^a-\mathcal{H}(\pi,p)$.  Let $\pi^a(\vec{x})$ and $p_a(\vec{x})$ be the static field configuration with a nontrivial winding number.  The quantum mechanics of the translational moduli can be obtained by substituting $\pi^a(\vec{x}-\vec{x}_0(t))$ and $p_a(\vec{x}-\vec{x}_0(t))$ to the Lagrangian and expanding it to the quadratic order.  We find, up to a total derivative term,
\begin{eqnarray}
L_{\text{eff}}&=&\int \mathrm{d}^dx\mathcal{L}_{\text{eff}}=(I/2)(y_0\dot{x}_0-x_0\dot{y}_0)+\cdots,\label{Ieff} 
\end{eqnarray}
with $I\equiv-\int\mathrm{d}^2x\epsilon^{ij}\nabla_ip_a\nabla_j\pi^a$.  In this effective theory, the momentum operator is given by $P^i=I\epsilon^{ij}x_0^j$ and the canonical commutation relation is $[x_0,y_0]=i\hbar I^{-1}$.  Thus, we have $[P^x,P^y]=I^2[y_0,-x_0]=i\hbar I$, and $I$ in Eq.~\eqref{Ieff} indeed represents the central extension $C^{xy}$.  As explained above, $-I$ represents the effective magnetic field $eB$ and the total flux $eBA=2\pi\hbar(n_0A)N$ is quantized to an integer in the unit of the flux quantum $2\pi\hbar$.

Let us discuss the relationship between the second cohomology $H^2(G/H)$ and the second homotopy $\pi_2(G/H)$.  When $\pi_1(G/H)=0$, $H^2(G/H)$ is isomorphic to $\pi_2(G/H)$ (Hurewicz theorem), and topologically nontrivial field configurations relevant for central extensions can be classified by $\pi_2(G/H)$.  For example, Skyrmions in the $\mathbb{C}\text{P}^n$ model fall into this category.  When $\pi_1(G/H)\neq0$, however, $\pi_2(G/H)$ may vanish even when $H^2(G/H)$ is nontrivial.  As an example, let us consider the case $G/H=T^2=\{(\theta_1,\theta_2)|\theta_i \in [0,2\pi)\}$.  The Lagrangian includes $n_0\hbar(\theta_1\dot{\theta}_2-\theta_2\dot{\theta}_1)/4\pi$
and $\omega=n_0\hbar\,\mathrm{d}\theta_1\wedge\mathrm{d}\theta_2/2\pi$ represents an element of $H^2(T^2)=\mathbb{Z}$.  As long as we impose $\theta^a(\vec{x},t)\rightarrow0$ as $|\vec{x}|\rightarrow\infty$, the winding number $N$ in Eq.~\eqref{general} vanishes as suggested by $\pi_2(T^2)=0$.  On the other hand, if we take $T^2$ as our spatial manifold and impose the periodic boundary conditions, $N$ measures the nontrivial winding of the map $\theta:T^2\rightarrow T^2$.  This example clarifies that it is $H^2(G/H)$, rather than $\pi_2(G/H)$, that is essential to central extensions.  Another interesting example is ${\cal M}=\mathbb{R}\text{P}^2  = S^2/{\mathbb Z}_2$ with $\pi_2({\cal M})={\mathbb Z}$, while $H^2({\cal M})=0$ because it is nonorientable.  It does not give rise to a central extension.

In fact, all of the above considerations hold even when the target space is a general K\"ahler manifold $K$ where the analog of a Skyrmion solution is a holomorphic map from the plane ${\mathbb C}\rightarrow K$.  We demonstrate this point in the Supplemental Material~\cite{SM}.

\paragraph{Central extension due to topological defects.}---
Finally, let us discuss another origin of central extensions: topological defects characterized by $\pi_1(G/H)$.  

The simplest example is the vortex in superfluid.  We take the Gross-Pitaevskii model~\cite{Pethick},
\begin{eqnarray}
\mathcal{L}=\hbar n\dot{\theta}-\frac{\hbar^2n}{2m}(\nabla\theta)^2-\frac{\hbar^2(\nabla n)^2}{8mn}-\frac{g}{2}(n-n_0)^2.\label{superfluid}
\end{eqnarray}
To regularize the following calculation, we subtract a surface term $\hbar n_0\dot{\theta}$, making the first term of the Lagrangian $\hbar(n-n_0)\dot{\theta}$.  We assume that the density $n(\vec{x},t)$ approaches to $n_0$ away from the vortex and vanishes at the core of the vortex, as required by the single-valuedness of $\psi=\sqrt{n}e^{-i\theta}$.  
The momentum operator is $P^i=\int\mathrm{d}^2x\,p\nabla_i\theta$, with $p=\hbar(n-n_0)$ and the commutation relation is
\begin{eqnarray}
&&(i\hbar)^{-1}[P^x,T^{0y}]=-\nabla_xT^{0y}+p\epsilon^{ij}\nabla_i\nabla_j\theta,\\
&&(i\hbar)^{-1}[P^x,P^y]=-2\pi n_0 N,
\end{eqnarray}
where we have used the relation $\epsilon^{ij}\nabla_i\nabla_j\theta(\vec{x},t)=2\pi N\delta^2(\vec{x})$.  This result is expected since each vortex has the charge $2\pi$ and feels the magnetic field $n_0$ in the dual description.  This commutation relation explains the well-known Magnus force $2\pi n_0N\hat{z}\times\dot{\vec{x}}_0$~\cite{Thouless} and the dispersion of the Kelvin wave~\cite{Kelvin}.  For a relativistic superfluid, the equilibrium density $n_0$ vanishes and thus $C^{xy}=0$.  Hence, two translational moduli are independent~\cite{KobayashiNitta}.

A meron texture in ferromagnets~\cite{Affleck,Fradkin} can be regarded as either half of a Skyrmion or a vortex in superfluids far apart from the center.  One can easily check that both pictures end up with the same extension.

A natural question to ask is whether other types of topological defects produce extensions. We explicitly checked that $\mathbb{Z}_2$ vortices for $\pi_1(\mathbb{R}\text{P}^2)={\mathbb Z}_2$ or $\pi_1(\mathbb{R}\text{P}^3)={\mathbb Z}_2$ do not.  This is expected since $C^{xy}$ has a ``sign" while the $\mathbb{Z}_2$ index does not.  In three spatial dimensions, we can also consider topological defects characterized by $\pi_2(G/H)$, {\it e.g.}\/, a monopole for $\pi_2(S^2)={\mathbb Z}$ or $\pi_2(\mathbb{R}\text{P}^2)={\mathbb Z}$. However, they do not produce nonzero extensions either, because they preserve three dimensional rotational symmetry.

There is one remaining subtle issue.  The superfluid Lagrangian in Eq.~\eqref{superfluid} has Galilean symmetry and thus it appears that the extension should not exist to be consistent with the Jacobi identities.  A similar problem arises when we consider a Skyrmion line configuration in $3+1$ dimensions.  If we take a box $[-L/2,L/2]^3$ as our spatial manifold and consider a Skyrmion texture in each $xy$ plane, we find the central extension $C^{xy}=4\pi (S/a^2) N (L/a)$.  In the large volume limit $L\rightarrow\infty$, however, the Lagrangian restores the three dimensional rotational symmetry and it appears that the extension should vanish according to the Jacobi identities.  Our resolution of this problem is the following: In the presence of a vortex or a Skyrmion line, the Galilean boost and spatial rotation are no longer well defined since they change the field configuration at the spatial boundary, and hence we do not need to consider their Jacobi identities. 

We thank Masahito Ueda for stimulating suggestions, and Igor Shovkovy, Uwe-Jens Wiese, Yoshimasa Hidaka, and other attendees of the workshop ``Effective Field Theory for Quantum Many Body Systems" at IFT UAM-CSIC (the Centro de Excelencia Severo Ochoa Program under Grant No. SEV-2012-0249) for fruitful comments.  We are grateful for useful discussions with Tomoya Hayata, Michikazu Kobayashi, Muneto Nitta, Yasuhiro Tada, Masaki Oshikawa, and Ashvin Vishwanath.  H. M. also thanks Scott Carnahan and Yoshitake Hashimoto for useful discussions on cohomology and representation theory.  We also thank Grigori Volovik for informing us of Ref.~\cite{Volovik}.  H. W. appreciates financial support from the Honjo International Scholarship Foundation.  The work of H. M. was supported by the U.S. DOE under Contract No.~DE-AC03-76SF00098, by the NSF under Grants No.~PHY-1002399 and No.~PHY-1316783, by the JSPS Grant No.~(C) 23540289, and by WPI, MEXT, Japan.

\bibliography{Refs}

\clearpage
\appendix
\onecolumngrid
\section{Supplemental Material\\for ``Noncommuting Momenta of Topological Solitons"}

Here we extend our analysis for homogeneous spaces $G/H$ to general K\"ahler manifold.

Suppose ${\cal M}$ is a K\"ahler manifold, whose metric $g_{a\bar{b}} = \partial_a \bar{\partial}_{\bar{b}} K$ is given by the K\"ahler potential $K(z^a, \bar{z}^{\bar{b}})$.  Then the K\"ahler form $\omega = i g_{a\bar{b}}\mathrm{d}z^a \wedge\mathrm{d}\bar{z}^{\bar{b}}$ is a nontrivial element of $H^2({\cal M})$.  The general form of the Lagrangian in $2+1$ dimension is 
\begin{equation}
\mathcal{L} = \frac{i}{2}\big(\bar{\partial}_{\bar{b}} K \dot{\bar{z}}^{\bar{b}}-\partial_a K \dot{z}^a\big)
+ g^T_{a\bar{b}} \dot{z}^a \dot{\bar{z}}^{\bar{b}}
- g^S_{a\bar{b}} \nabla_i{z}^a \nabla_i{\bar{z}}^{\bar{b}}\ .
\end{equation}
Here we made it clear that the K\"ahler metrics $g^T_{a\bar{b}}=\partial_a \bar{\partial}_{\bar{b}} K^T$, $g^S_{a\bar{b}}=\partial_a \bar{\partial}_{\bar{b}} K^S$ may in general be different from $g_{a\bar{b}}$ obtained from $K$. It is convenient to introduce a complex notation for the spatial coordinates $w = x+iy$, $\nabla = \frac{1}{2} (\nabla_x - i \nabla_y)$.  The energy $E$ of a static field configuration is given by
\begin{eqnarray}
E&=& \int d^2 x\,g^S_{a\bar{b}} \nabla_i{z}^a \nabla_i{\bar{z}}^{\bar{b}}\nonumber\\
&=& \int d^2 x\,2g^S_{a\bar{b}}\left[\left(\nabla z^a \bar{\nabla}{\bar{z}}^{\bar{b}} - \bar{\nabla}{z}^a \nabla\bar{z}^{\bar{b}}\right)+2 \bar{\nabla}{z}^a \nabla\bar{z}^{\bar{b}} \right] \nonumber \\
&\geq& \int d^2 x\,ig^S_{a\bar{b}}\epsilon^{ij}\nabla_iz^a \nabla_j\bar{z}^{\bar{b}} 
= 2\pi\hbar n_0 N^S\ .
\end{eqnarray}
Here we used the fact that $g^S_{a\bar{b}}$ is positive definite to derive the inequality.  The last expression is nothing but the pull-back of the K\"ahler form $\omega^S = i g_{a\bar{b}}^S\mathrm{d}z^a \wedge\mathrm{d}\bar{z}^{\bar{b}}$.  To minimize the energy for a fixed $N^S>0$, we need $\bar{\nabla}{z}^a = 0$, namely $z^a(w)$ is a holomorphic map ${\mathbb C}\rightarrow {\cal M}$ (anti-holomorphic for $N^S<0$).  Expanding the solution for its translational moduli, $z^a = z^a(w-w_0(t))$ with $w_0=x_0+iy_0$, the first term in the Lagrangian becomes
\begin{eqnarray}
\lefteqn{
\int d^2 x\,\frac{i}{2}\big(\bar{\partial}_{\bar{b}} K \dot{\bar{z}}^{\bar{b}}-\partial_a K \dot{z}^a\big)}\nonumber \\
&=&\left[\int d^2 x\,2g_{a\bar{b}}(\nabla z^a\bar{\nabla}\bar{z}^{\bar{b}}-\bar{\nabla}z^a\nabla\bar{z}^{\bar{b}})\right]\frac{1}{4i}(\bar{w}_0\dot{w}_0 - \dot{\bar{w}}_0w_0) \nonumber \\
&=&\left[\int d^2 x\,ig_{a\bar{b}}\epsilon^{ij}\nabla_iz^a\nabla_j\bar{z}^{\bar{b}}\right]\frac{1}{2}\epsilon_{ij} x_0^i \dot{x}_0^j \nonumber \\
&=&(2\pi \hbar n_0 N) \frac{1}{2} \epsilon_{ij} x_0^i \dot{x}_0^j,
\end{eqnarray}
for a nontrivial K\"ahler class.  The skyrmion for a ferromagnet is a special case of this general consideration.

We believe the analysis would extend to even wider class of target manifolds as long as they permit presymplectic structure and have a nontrivial $H^2$.

\end{document}